\documentstyle[12pt]{article}

\title{Exchange statistics in 1D: from the viewpoint of exact
solution}
\author{K.N. Ilinski$^{1,2}$,  G.V.Kalinin$^{3}$, V.S.Kapitonov $^{4}$
\\
{\small\it $^{1}$ School of Physics and Space Research, University of
Birmingham,}
\\
{\small\it Birmingham B15 2TT, United Kingdom.}
\\
{\small\it $^{2}$ Institute of Spectroscopy, Russian Academy of Sciences,}
\\
{\small\it Troitsk, Moscow region, 142092, Russian Federation.}
\\
{\small\it $^{3}$ Institute of
Physics, Physics Department of St-Petersburg University,}
\\
{\small\it St-Petersburg, 198904, Russian Federation.}
\\
{\small\it $^{4}$ St-Petersburg
Technological University, Moskovskii prosp.  26,}
\\
{\small\it St-Petersburg, 198013, Russian Federation.}}
\date{ }

\begin{document}

\setcounter{page}{0}
\maketitle
\thispagestyle{empty}

\begin{abstract}
We show
that the exchange statistics have consequences in 1D systems with
compact topology, contrary to the common opinion
that exchange statistics is arbitrary in 1D. As examples of
non-trivial statistical behavior we exactly calculate the partition function
and correlators for  systems of free q-particles on compactified
chains using functional integral techniques and the supersymmetric trick. In
particular we consider a spin 1/2 XY-chain
with periodic boundary conditions that corresponds to the case of q=-1.

\end{abstract}
\newpage

Last decade there has been considerable interest in various deformations
of quantum statistics which interpolate between the Bose- and Fermi- cases.
Such deformations can be separeted into
the deformations of
exchange phase which appears under the permutation of particles (exchange
statistics with braiding \cite{W1} or without it \cite{I}) and
the deformations of Pauli principle
(so called exclusion statistics \cite{Hald}). While the last can be defined
in any space dimension the possibility to introduce non-trivial exchange
statistics crucially depends on the dimension.

Moreover there is a common opinion that the exchange statistics is
something arbitrary in 1D. It is due to the fact that in 1D for
the particles with hard-core condition on a finite interval there no
real permutation is allowed and, hence, the exchange phase never
plays a role. The situation changes if the interval is reduced to a
circle with periodic boundary conditions. This leads to a
non-trivial loop which gives new possibilities to permute particles via
the `glued` boundary. Then the effects of the exchange statistics immediately
exhibit themselves as we will demonstrate in this Letter. However we should add
that we deal with lattice systems because the effects of the loops are
negligible in continuous limite.

The present calculations use the functional integral
formalism.  This is not surprising if we remember that the formal form of
the functional integral representation for the thermodynamical quantities are
essentially the same for both Fermi- and Bose- statistics and the
difference is reflected only in the nature of fields and boundary
conditions \cite{Popov}. We will now show that the
statistics can be taking into
account by changing the boundary conditions both in imaginary
time and space which lead to the transmutation of distribution functions.
In the framework of our model we calculate exactly such distribution
functions and, moreover, the result is simple. This hints at the
form of multidimensional analogues (see discussion after Eq.(\ref{n})).

The point is illustrated by the example of free q-fermions, i.e. the
particles with deformed exchange statistics defined by the algebra of
commutation relations
for the creation (annihilation) operators $a^+_i,(a_i)$ at
the $i$-th site of the compact chain with length $M$. The
relations are split into two parts: the first
describes the fermionic algebra on each site:
\begin{equation}
a^{+}_i a_i + a_i a^{+}_i=1 \ , \qquad a_i^{2}=0  \ ,\qquad
(a_i)^{+}=a^{+}_i \ ,
\label{fermi}
\end{equation}
and the second  gives  the
exchange statistics with  deformation parameter
$q=e^{2\pi ir/n}, n \geq r\geq 1$:
\begin{equation}
a_i a_j = -q a_j a_i \ , \quad  a^{+}_i a_j = -q^{-1} a_j a_i^{+} \ ,
\label{exchange}
\end{equation}
where the ordering $1 \leq i < j \leq M $ is assumed.

Presently we will consider a gas of free q-fermions with the Hamiltonian:
\begin{equation}
H=\sum_{i=2}^{M}
(A_{i,i-1}a^{+}_{i}a_{i-1}+1/2B_{i}a^{+}_{i}a_{i})+qA_{1M}a^{+}_{1}a_{M} +
1/2B_{1}a^{+}_{1}a_{1}+H.c.
\label{H}
\end{equation}
One of the main purposes of the Letter is to examine the
consequences of $q$-deformed exchange statistics for the thermodynamical
quantities on the basis of the Hamiltonian (\ref{H}). It turns out that the
partition function and correlators for the model can be calculated exactly.
So we are going to do for the exchange statistics the same that was done
by Ha \cite{Ha} for the exclusion statistics: to examine statistics from
the point of view of exact solutions. It should shed light on the
whole problem.

First of all let us describe  and discuss the main results for
the thermodynamical quantities associated with the Hamiltonian (\ref{H})
and then give the outline of the calculations.
The starting point for this is the exact expression for the partition
function of the system, $Z=\mbox{Tr\thinspace } \exp (-\beta(H-\mu\hat N))$,
at temperature $1/\beta$ and chemical potential $\mu$, where
$\hat N=\sum^M_{i=1}a^{+}_{i}a_{i}$ is the number operator for
q-particles:
\begin{equation}
Z=\frac{1}{n}
\sum_{l=0}^{n-1}\sum_{p=0}^{n-1} q^{-pl}
\prod_{k^{m}_{l}}
(1 + q^{p} e^{\beta (\mu - \epsilon^{l}(k^{m}_{l}))})
\label{Zq}
\end{equation}
Here the notation $\epsilon^{l}(k^{m}_l)$ was introduced for the
$k^{m}_{l}$-th eigenvalue of the hopping matrix $||A_{ij}||$ on the circle
with $q^{l}$-periodic boundary conditions (which corresponds to the
substitution of $q^{l}A_{1M}$ instead of $A_{1M}$ in the hopping matrix).
For the homogeneous chain ($A_{ij}=\mbox{\cal A}$) these eigenvalues have
the form:
\begin{equation}
\epsilon^{l}(k^{m}_l)=B+2A\cos (k^{m}_l)\ , \qquad
k^{m}_l=\frac{2\pi}{M}(m-lr/n)
\label{epsilon}
\end{equation}
with $m=0,\ldots ,M-1$.
Using  expression (\ref{Zq}) we immediately obtain the representation
for the distribution function $n(\beta, \mu)$ of the q-particles in terms
of transmuted Fermi-Dirac (or Bose-Einstein) functions
$f_{q}(x)=1/(1+q^{-1}\exp(x))$:
\begin{equation}
n(\beta, \mu)=\frac{
\sum_{l=0}^{n-1}\sum_{p=0}^{n-1} q^{-pl} \sum_{k^{m}_{l}}
f_{q^{p}}(\beta(\epsilon^{l}(k^{m}_{l})-\mu))
\prod_{k^{m}_{l}}
(1 + q^{p} e^{\beta (\mu - \epsilon^{l}(k^{m}_{l}))})}{
M \sum_{l=0}^{n-1}\sum_{p=0}^{n-1} q^{-pl}
\prod_{k^{m}_{l}}
(1 + q^{p} e^{\beta (\mu - \epsilon^{l}(k^{m}_{l}))})} \ .
\label{n}
\end{equation}
So we see that in the Eqn.(\ref{n}) q-deformed Fermi-Dirac
functions, with all allowed powers of the deformation parameter $q$, make
contributions. Such
functions naturally interpolate between fermionic distribution function
($q=1$) and bosonic distribution function ($q=-1$).  Moreover they pick up
all $q$-periodic boundary conditions which also reflects $q$-exchange
factor gained due to the permutation of particles via the loop.

The note about all  $q$-periodic boundary conditions
 let us to look at the problem from the different
position. Indeed, from the original form of the Hamiltonian
(\ref{H}), which is quadratic on the creation (annihilation) operators, it
is not difficult to see that, using the appropriate diagrammic technique
(for example in the $q$-field technique \cite{IS}), all contributions to the
thermodynamical quantities will be given by loop diagrams with some
exchange factors. And this does not depend on the space dimension. On the
other hand, the examination of the expression (\ref{Zq}) for the 1D case
may prompt an ansatz for the partition function in $n$D after the summation
of such contributions for different loops: the sums over all loops in the
configuration space, with all allowed $q$-periodic boundary conditions for
them and all possible $q$-deformations of Fermi-Dirac constructions.

Let us turn our attention to the particular case $q=-1$. In this case the
Hamiltonian (\ref{H}) can be realized as a Hamiltonian of the spin $1/2$
compact XX-chain in a magnetic field $B$ (for the sake of simplicity we
deal with the homogeneous case):
\begin{equation}
\begin{array}{ll}
H = \sum_{i=2}^{M} & (A_x s^{x}_{i} s^{x}_{j} + A_y s^{y}_{i} s^{y}_{j}
-Bs^{z}_{i})- B_{1}s^{z}_{1}\\
 & + A_x s^{x}_{1} s^{x}_{M}+ A_y s^{y}_{1} s^{y}_{M} +MB/2
\end{array}
\label{SpinHam}
\end{equation}
if we identify $A_x=A_y$ and the creation (annihilation) operators with
spin upper (lower) operators are usual:
\begin{equation}
a^{+}_{i} = s^{+}_{i} = s^{x}_{i} - i s^{y}_{i} \ , \quad
a_{i} = s^{-}_{i} = s^{x}_{i} + i s^{y}_{i} \ .
\nonumber
\end{equation}
After the introduction of anisotropy $A_{x(y)}=2(\mbox{\cal A} \pm \gamma)$
Hamiltonian (\ref{SpinHam}) becomes the Hamiltonian of a spin
$1/2$ compact XY-chain in the magnetic field $B$. As we will show below, we may
obtain the exact formulae for
the thermodynamical quantities for the Hamiltonian of XY-model
in compact lattice topology (in contrast to previous works where such
quantities were calculated in thermodynamical limite \cite{SML,N,BM,Iz1,Iz2}).
Now we only
state them. The partition function of the model $Z(\beta,B)$ contains
four terms, two of which have fermionic nature and the others  have
bosonic ones:
\begin{equation}
Z = \frac{1}{2}(Z_{f}^{+} + Z_{f}^{-} + 1/Z_{b}^{+} - 1/Z^{-}_{b}) \ .
\label{ZB}
\end{equation}
Here $Z_{f}^{\pm}(Z_{b}^{\pm})$ are fermionic (bosonic) partition functions
$$
Z_{f}^{\pm} = e^{-\beta BM/2}\prod_{k^{\pm}} e^{\beta E(k^{\pm})/2}
(1+e^{-\beta E(k^{\pm})})
$$ $$
\quad Z_{b}^{\pm} = e^{\beta BM/2}
\prod_{k^{\pm}} e^{-\beta E(k^{\pm})/2} /(1-e^{-\beta E(k^{\pm})})
$$
for systems with the energy spectra
$$
E(k^{\pm}) = (B+2\mbox{\cal A}\cos(k^{\pm}))
\sqrt{1 + \frac{(2\gamma\sin(k^\pm))^2}{(B+2\mbox{\cal A}\cos(k^\pm))^2}} $$
and antiperiodic (periodic) boundary conditions:
$$
k^{+}=\frac{2\pi}{M} (m+1/2)\ , \quad k^{-}=\frac{2\pi}{M}m  \quad
m=0,\ldots ,M-1 \ .
$$
We also calculated correlators for various spin components. Since
the results are very cumbersome, they will be reported elsewhere.

Now let us outline the calculational procedure which leads to
expressions (\ref{Zq}),(\ref{n}) and then give the
modification for the case of XY-model (\ref{SpinHam}).
To this end we apply to the creation (annihilation) operators $a^+_i,(a_i)$
some analog of the Jordan-Wigner transformation just as it was performed
in 1961 by E.Lieb, D.Mattis and T.Schults in
Ref.\cite{SML} for the case of the Paulion chain (q=-1):
\begin{equation}
a_i = q^{\sum^i_{k=1}c^+_kc_k} c_i \ ,\qquad
a^+_i = c^+_i q^{-\sum^i_{k=1}c^+_kc_k}\ .
\label{JW}
\end{equation}
Where the above operators $c^+_i,(c_i)$ are respectively creation
(annihilation) operators of spinless fermions at the $i$-th site.
The operator (\ref{H}) is cast  in the form:
\begin{equation}
H=\sum_{i=2}^{M}(A_{i,i-1}c^{+}_{i}c_{i-1}+1/2B_{i}c^{+}_{i}c_{i})+
A_{1M}q^{\hat N}c^{+}_{1}c_{M} + 1/2B_{1}c^{+}_{1}c_{1} + H.c.
\label{Hq}
\end{equation}
where $\hat N=\sum^M_{i=1}c^+_i c_i$ is the number operator of the fermions.
To proceed we introduce the set of operators $\{P_{l}\}^{n-1}_{l=0}$  which
are projection operators on the subspaces of $l(\mbox{mod\thinspace } n)$
particles:
\begin{equation}
P_{l} = \frac{1}{n}
\sum_{p=0}^{n-1} q^{-lp}q^{\hat Np}\ .
\label{P}
\end{equation}
Making use of these projection operators
with the obvious property $\sum^{n-1}_{l=0}P_{l}=I$ we then obtain the
following representation for the partition function
$Z=\mbox{Tr\thinspace } \exp (-\beta(H-\mu\hat N))$:
\begin{equation}
Z=\sum^{n-1}_{l=0}\mbox{Tr\thinspace } \exp (-\beta(H-\mu\hat N)) P_{l}
\label{sum}
\end{equation}
that allows to replace in each term of the sum (\ref{sum}) the multiplier
$q^{N}$ in the Hamiltonian (\ref{Hq}) by the  numerical constant $q^{l}$.
Moreover due to the commutation of the operators $H$ and $\hat N$ each
$p$-th term of the expression (\ref{P}) for the projection operators can be
taken into account by a shift of the chemical potential $\mu$ to the value
$\mu^{(p)}=\mu+2\pi i p r/n\beta$. This leads to the representation for
the partition function:
\begin{equation}
Z=\frac{1}{n}
\sum_{l=0}^{n-1}\sum^{n-1}_{p=0} q^{-lp}\mbox{Tr\thinspace } \exp
(-\beta(H^{(l)}-\mu^{(p)}\hat N))
\label{ZP}
\end{equation}
where $H^{(l)}$ are
$q^{l}$-periodical square fermionic Hamiltonians. After the making of use
of the standard results for the partition function of a free fermion system we
regain to the Eq.(\ref{Zq}). In the same way various
averages and correlation functions can be obtained. For example:
$$
\langle a_{1}(\tau) a^{+}_{L+1}(0)\rangle = \frac{e^{-\tau MB}}{nZ}
\sum_{l=0}^{n-1}\sum^{n-1}_{p=0} q^{-lp} \mbox{Det\thinspace}C^{l}_{p}(\tau)
(C^{l}_{p}(\tau))^{-1}_{1,L+1}
$$
where the matrix $C$ is defined by the relation
\begin{equation}
\begin{array}{ll}
C^{l}_{p}(\tau)_{m^{\prime},m^{\prime \prime}} = &
\frac{1}{M} \sum_{k^{m}_{l+1}} e^{\tau(\epsilon^l(k^{m}_{l+1})-\mu)+i
k^{m}_{l+1} (m^{\prime}-m^{\prime \prime})} \\
 &
+\frac{q^{p}}{M} \sum_{k^{m}_{l}} e^{(\tau-\beta) (\epsilon^l(k^{m}_{l})-\mu)
+ik^m_l (m^{\prime}-m^{\prime \prime})} (1 + (q^{-1}-1)
\theta (L-m^\prime))
\end{array}
\end{equation}
(here we used lattice $\theta $ symbol which is continuous from right).
It is not difficult to check that this average reproduces the formula
(\ref{n}) for the distribution function if $\tau , L \rightarrow 0$.

By the same way we calculate the generating functional
$\langle e^{\alpha Q(L)} \rangle$ with $Q(L) = \sum^L_{k=1} c^+_k c_k$.
Using the expression (\ref{Zq}) for the partition function $Z$ we have
\begin{equation}
\begin{array}{l}
\langle e^{\alpha Q(L)} \rangle =
\frac {e^{\beta MB/2}}{nZ} \sum_{l=0}^{n-1} \sum_{p=0}^{n-1} q^{-lp}\\
\mbox{Det\thinspace} \left(
\delta_{k^m_l,k^{m^\prime}_l} + \frac{e^{\alpha} - 1}{M}
\frac{1}{1 + q^{-p}  e^{\beta(\epsilon^l(k^m_l)-\mu)}}
\frac{\sin\frac L2 (k^m_l - k^{m^\prime}_l)}
{\sin\frac 12 (k^m_l - k^{m^\prime}_l)}
\right)
\prod\limits_{k^m_l}
\left( 1 + q^p e^{\beta(\mu-\epsilon^l(k^m_l))} \right)
\end{array}
\end{equation}

Up to this point we did not use the functional integral approach.
However it allows us to look at the problem from another viewpoint and,
furthermore, is closely connected with the modification of the calculations
for the case of XY-chain.  Indeed, each term of the sum (\ref{ZP}) can be
represented in the form of a function integral \cite{Popov}:
$$
\mbox{Tr\thinspace } \exp
(-\beta(H^{(l)}-\mu^{(k)} N)) = \int D\bar{\xi}(\tau) D\xi (\tau)e^{S_{lp}}
$$
over the Grassmann fields $\bar{\xi}(\tau),{\xi}(\tau)$ with antiperiodic
boundary conditions $\bar{\xi} (\beta) = -\bar{\xi} (0)$, ${{\xi}(\beta)}
= -{{\xi}(0)}$ and the action
\begin{equation}
\begin{array}{ll}
S_{lp} = &
\int_{0}^{\beta} d\tau \{\bar{\xi}\frac{\partial \xi}{\partial \tau} -
H^{(l)}(\bar{\xi},\xi) + \mu \bar{\xi}\xi \} + \int_{0}^{\beta} d\tau
 \frac{2\pi i p r}{n \beta}\bar{\xi}\xi \\ & = S_{l} + \delta S_{p}
\end{array}
\end{equation}
Under the change of variables
$\bar{\xi}(\tau) \rightarrow \bar{\xi}(\tau)e^{-i\frac{2\pi p r}{n
\beta}\tau}$, ${\xi}(\tau) \rightarrow {\xi}(\tau)e^{i\frac{2\pi p r}{n
\beta}\tau}$ the last term of the previous expression, $\delta S_{p}$,
disappears, that, however, is compensated by the changing of the boundary
conditions for the fields of integration:
\begin{equation}
\bar{\xi} (\beta) =
-q^{-p}\bar{\xi} (0) \ , \quad {\xi}(\beta) = -q^{p}{\xi}(0) \ , \label{BC}
\end{equation}
(a procedure of such type was used in Ref. \cite{PF} to construct
the Feynman diagram technique for spin systems). As a result we reach the
following functional integral representation of the partition function:
\begin{equation}
Z=\sum_{p=0}^{n-1}\sum^{n-1}_{l=0} q^{-lp} \int D\bar{\xi}(\tau) D\xi
(\tau)e^{S_{l}} / n
\label{ZFI}
\end{equation}
provided with the $q^{p}$-periodic
boundary conditions (\ref{BC}). In this form the equation turns out to be
valid for the case of XY-model for spin 1/2, although  the
justification needs to attract other ideas.

Let's clarify the problem arising in the case of XY-model.
Indeed, after the
Jordan-Wigner transformation (\ref{JW}) the Hamiltonian (\ref{SpinHam})
becomes  the operator:
\begin{equation}
\begin{array}{ll}
H= & \sum_{i=2}^{M}(\mbox{\cal A}c^{+}_{i}c_{i-1}+1/2B_{i}c^{+}_{i}c_{i} +
\gamma c^{+}_{i}c^{+}_{i-1}) +1/2B_{1}c^{+}_{1}c_{1}\\
 & -\mbox{\cal A}(-1)^{\hat N}c^{+}_{1}c_{M} -
\gamma (-1)^{\hat N}c^{+}_{1}c^{+}_{M} +H.c.
\end{array}
\label{H12}
\end{equation}
which differs from the Hamiltonian (\ref{Hq}) by
additional terms with the anisotropy parameter $\gamma$. They cause the
Hamiltonian to no longer commute with the number operator for
particles $\hat N$, that does not allow us to use device of  shifting the
chemical potential.  However the Hamiltonian still commutes with the
operator $\tau=(-1)^{\hat N}$. This leads to the formula, which is
analogous to the Eqn.(\ref{ZP}):
\begin{equation}
\begin{array}{ll}
Z= \frac{1}{2}( & \mbox{Tr\thinspace } e^{-\beta(H^{(+)}-\mu\hat N)}+
\mbox{Tr\thinspace }
e^{-\beta(H^{(-)}-\mu\hat N)}+ \\
 & \mbox{Tr\thinspace } \tau e^{-\beta(H^{(+)}-\mu\hat N)}- \mbox{Tr\thinspace
}
\tau e^{-\beta(H^{(-)}-\mu\hat N)})
\end{array}
\end{equation}
where $H^{(\pm)}$ coincide with the Hamiltonian (\ref{H12}) after the
substitution $\tau=(-1)^{\hat N}\rightarrow \pm 1$. Although
the two last terms cannot be calculated by changing of chemical
potential, taking into account $\tau$ in them is also achieved by
changing of boundary conditions (\ref{BC}). This was demonstrated in the Ref.
\cite{CG} into the framework of supersymmetric quantum mechanics, where
the operator $\tau$ and $\mbox{Tr\thinspace } \tau e^{-\beta H}$
play the role of supersymmetric involution and supersymmetric Witten index
of the Hamiltonian $H$. That is why we term this the {\it supersymmetric
trick}.

All this returns us to Eq.(\ref{ZFI}) with $q=-1$, which after the
use of the Bogoliubov transformation for the variables of integration leads to
the formula (\ref{ZB}). We want to add that the same trick was used to
calculate correlation functions.

The Letter concentrates essentially on the principal questions concerning
exchange statistics. However the formulae obtained in the Letter can have
various physical applications.
Let's briefly mention them.
They describe the compact finite chain
in contrast to the classical papers \cite{SML,N,BM} and more
recent ones \cite{Iz1,Iz2} where thermodynamical quantities were
calculated in the thermodynamical limit. In this limit the
statistical effects, which are proportional to the inverse size of the
system, $1/M$, are negligible and the formulae of the paper are transformed
into the known results. However
recently  a lot of attention was attracted by the theory
of so-called J-aggregates, i.e. molecular aggregates with an unusually sharp
absorption band (\cite{J},\cite{KS} and Refs. wherein) where the advances
were connected with the use of exact results on 1D chains. Frenkel
excitons in such long molecules obey the Paulionic commutation
relations \cite{A0}, or, in a more general case (if we will take into account
retardation effects \cite{PS}), q-particle commutation relations. We hope
that theformulae  obtained  will be used in the theory of the nonlinear
response
of long cyclic molecules where a thermodynamical limit is not appropirate.
On the other hand XY-model recently arose in  models of
adsorption processes with diffusional relaxation \cite{St}.
Here the forlulae of this Letter can also find applications to
describe the compact case. The same one can be said about the application of
results to the theory of defectons where the statictics of
defectons is exactly Paulionic statistics\cite{Pushk}.

In conclusion, we exactly calculated the partition functions
and correlators for systems of free particles with q-deformed
exchange statistics and the spin 1/2 XY-model on a compact chain with periodic
boundary conditions.
We demonstrated that the deformation of the statistics can be taken into
account by changing the boundary conditions for the
fields of integration in the functional integral framework that lead to
the transmutation of distribution functions. In such a way deformed
Fermi-Dirac functions appeared. We also stated a conjecture about the form of
thermodynamical quantities in higher dimensions.

We wish to thank V.M.Agranovich, J.M.F.Gunn,  A.G.Izergin,
I.V.Lerner, A.S.Stepa\-nen\-ko for useful discussions.  This work was
supported (K.I.,V.K.) by the International Science Foundation, Grant N
R4T000, the Russian Fund of Fundamental Investigations, Grants N 94-02-03712
and
N 95-01-00548
and partially (K.I.) by the UK EPSRC under Grant number GR/J35221, Euler
stipend of German Mathematical Society and grant INTAS-939.

\end{document}